# OPTICAL IDENTIFICATION OF FEW-LAYER GRAPHENE OBTAINED BY FRICTION

Manuel Mailian[1], Kamran ul Hasan[2,3], Aram Mailian[4*]

[1] LTX-Credence Armenia, 2 Adonts str., 0014, Yerevan, Armenia.

[2] Department of Science and Technology (ITN), Linköping University, Campus Norrköping, SE 60 174 Norrköping, Sweden.

[3] Centres of Excellence in Science & Applied Technologies (CESAT) Islamabad, Pakistan.

[4] Institute for Informatics, 1 P. Sevak str., 0014, Yerevan, Armenia.

A friction based method is proposed for obtaining graphite layers. Freestanding thin structures of graphite containing several layers of graphene are obtained by rubbing a graphite rod on the surfaces of NaCl substrate and then dissolving the substrate in water. An optically transparent and mechanically strong lamina of few-layer graphene is found in the obtained structures.

Since the Scotch tape separation of single-layer graphite crystals [1], different approaches of straightforward separation were worked out for obtaining graphene. In this letter we report on a new technique of separating graphite structures obtained by rubbing graphite bulk on water soluble NaCl substrate.

We performed *en route* microscopic observation of the developed structure. Optical images were taken in transmission and reflection modes under the illumination of non-polarized white light. For the side illumination, the angle between the light and the structure plane was $\sim 20^{\circ}$.

The substrate of NaCl crystal was prepared by rubbing its surface, along a straight line, with 30 μm abrasive grading sandpaper, Figure 1a. Then the substrate was observed in both transmission and reflection modes. The sandpaper dug trenches on the surface of the substrate were seen as a set of alternating parallel strips, Figure 1b.

---

[*] Corresponding author. Tel: +374 94 100-391. E-mail: amailian@ipia.sci.am

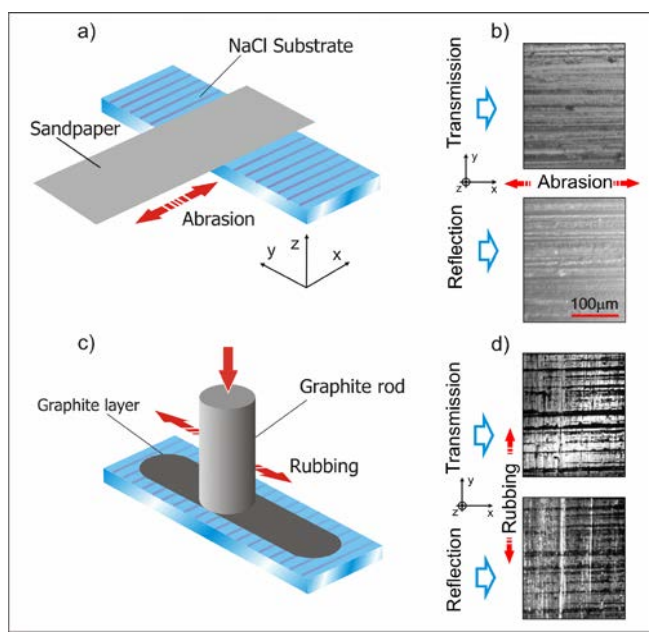

Figure 1: Obtaining graphite layers on a NaCl substrate. a) Processing of the surface of a NaCl crystal. Red arrows show the directions of sandpaper abrasion. b) Images of the surface of processed crystal. c) Development of a graphite structure on the surface of processed NaCl substrate. Vertical red arrow shows the direction of pressing on graphite rod and arrows parallel to the substrate surface show the direction of rubbing. The graphite rod was held vertical to the substrate surface during the whole cycle of rubbing. d) Images of developed graphite structures.

Graphite structures were developed by manifold (up to 60) layer-on-layer rubbing of a graphite rod against the prepared substrate surface, Figure 1c. Pressure of ~ 5 MPa was applied on graphite rod throughout the rubbing process.

Due to rubbing, two sets of dark strips appeared in transmission images; one along the sandpaper dug trenches and the next, along the rubbing path, Figure 1d - upper panel. Vast areas between the dark strips remain bright, i.e. visibly transparent throughout the entire rubbing cycles.

In reflection images, the dark strips along the graphite rubbing path turned into shining streaks, Figure 1d - lower panel. Between these shining streaks one could see darker and flat fields which were observed as bright areas in the corresponding transmission image, Figure 1d – upper panel.

Freestanding structures were obtained by employing the following technique, Figure 2.

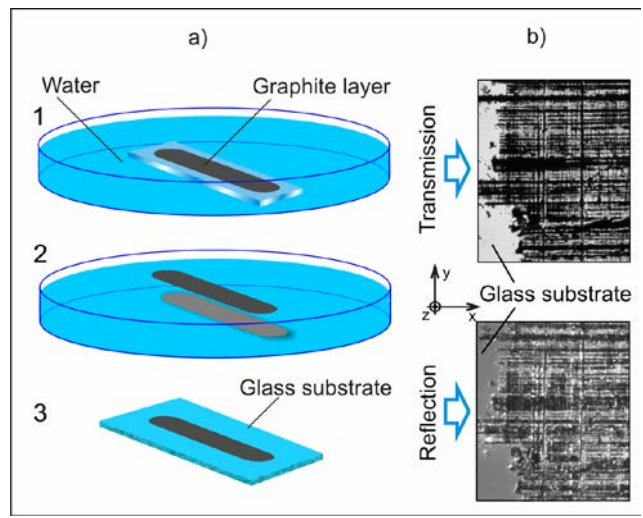

Figure 2: Separation and optical images of obtained structure. a) Graphite structure rubbed on NaCl substrate is immersed into water, then, due to dissolution of the substrate it detached from the substrate. Then, the floating structure is transferred on a glass substrate. b) Images of separated graphite structures, positioned face-up.

The NaCl crystal with rubbed off graphite coating was immersed into water (step 1→2). With the dissolution of the substrate, the coating was detached and floated in water. Then, the structure was transferred onto a glass substrate (step 2→3) in either of two positions: rubbed face up, or rubbed face down. The images of structures transferred rubbed face up repeat the texture of the structures developed on substrate, Figure 2b. The strips shining in reflection mode (Figure 1d) also shine in transferred structures (Figure 2b).

Both in the cases of rubbed and face-up transferred structures, Figures 1d and 2b, strips not shining in reflection mode, are the projections of graphite filled trenches at the structure bottom, whereas the perpendicular shining streaks (along y-axis) are the traces of reflections from the surface created during friction. Between them, we deduce that a binding transparent graphite layer is created.

To estimate the thickness of transparent areas, we calculated the through-thickness optical transmission in the vicinity of a randomly chosen check-like segment containing a bright area, Figure 3. With the assumption that a monolayer of graphite absorbs 2.3 % of the incident light in visible region [2,3], calculations yielded that the bright transparent areas are several monolayer thin. Quantitatively, for a randomly chosen segment of ~25,000 μm², 2 to 6 single layer thick areas cover more than 20 % of the surface of given sample, Figure 3c. 8.8 % of the studied area is of 2 single layers, 7 % - 4 single layer and 5 % - 6 single layer thick.

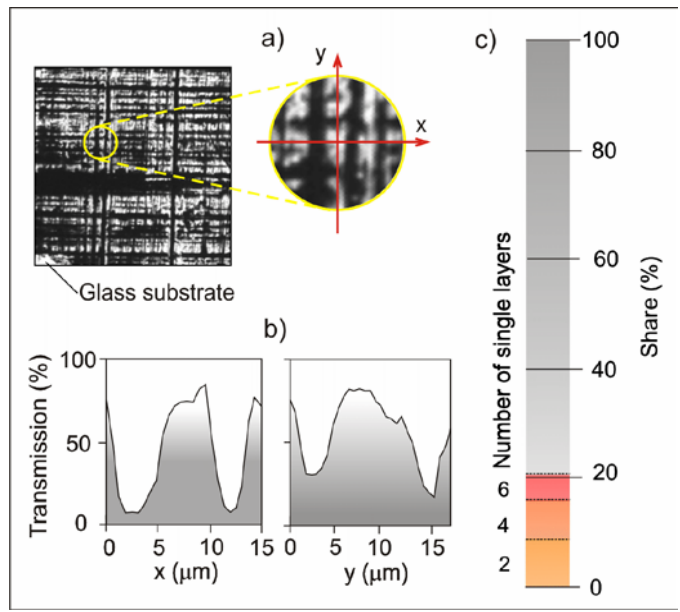

Figure 3: a) A check-like segment of a structure image (marked in yellow ring); the structure was transferred onto a glass substrate, face up. b) The profile of optical transmission along x and y lines (bellow). c) The bar graph shows the share of transparent layers obtained for ~20.000 μm² area of structure.

These data can allow one to sketch out the through-thickness configuration of obtained structures, Figure 4. Based on the following two observations: a) the divergence in shining behavior under side-illumination b) the presence of flat bright areas between the dark strips in transmission images, we deduce that we captured three substructures: i) a set of ridge-like structures at the bottom, formed due to filling of sandpaper abraded trenches, ii) another set of ridge-like structures at the top, formed due to rubbing and iii) a supporting transparent and strong lamina confined in-between the sets of ridge-like structures.

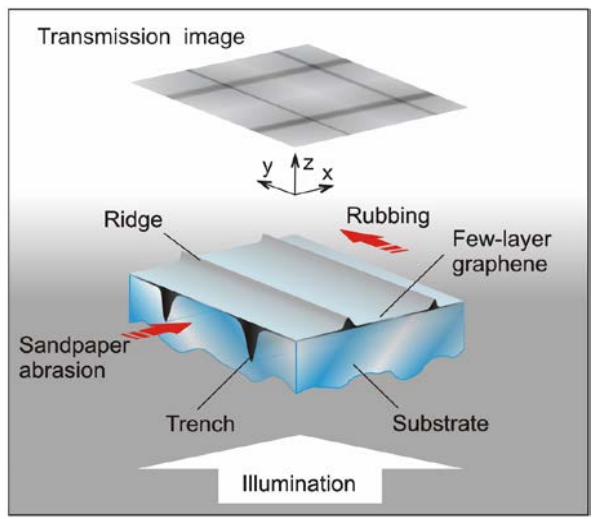

Figure 4: Schematic presentation of the structure in the vicinity of the check-like item.

In conclusion, rubbed off graphite structures were transferred onto glass substrate, and then observed in transmission and reflection mode. Analysis of the received optical images identified a transparent lamina of few-layer graphene in these structures. We explain its origin in the following way: The repetitive action of friction forces i.e. combined action of cleaving → transferring → sliding → pressing of graphite flakes is most likely responsible for the development of a self-organized structure [4, 5] in the form of optically transparent graphite layers with the thickness of few basic carbon planes. This simple yet productive method of obtaining carbon layers using soluble substrates holds promise for future applications.

**References**

1. Novoselov KS, Geim AK, Morozov SV, Jiang D, Zhang Y, Dubonos SV, Grigorieva IV, Firsov AA. Electric Field Effect in Atomically Thin Carbon Films. Science 2004, 306, 666-669.
2. Mak KF, Sfeir MY, Wu Y, Lui CH, Misewich JA, Heinz TF. Measurement of the Optical Conductivity of Graphene. Physical Review Letters 2008; 101(19): 196405.
3. Nair RR, Blake P, Grigorenko AN, Novoselov KS, Booth TJ, Stauber T, et al. Fine Structure Constant Defines Visual Transparency of Graphene. Science 2008; 320(5881): 1308.
4. Bershadskii LI. Self-organization of Tribosystems and Conception of Wear Resistant. Journal of Friction and Wear 1992;6(13): 1077–94.
5. Self-Organization During Friction: Advanced Surface-Engineered Materials and Systems Design: CRC Press; 2006.